\documentclass[twocolumn,amsmath,amssymb,prl,groupedaddress,nobibnotes,showpacs]{revtex4}
\usepackage{graphicx}
\usepackage{dsfont}

\begin{document}
\title{Reversing the Weak Quantum Measurement for a Photonic Qubit}

\author{Yong-Su Kim}\email{yskim25@postech.ac.kr}
\affiliation{Department of Physics, Pohang University of Science and Technology (POSTECH), Pohang, 790-784, Korea}

\author{Young-Wook Cho}
\affiliation{Department of Physics, Pohang University of Science and Technology (POSTECH), Pohang, 790-784, Korea}

\author{Young-Sik Ra}
\affiliation{Department of Physics, Pohang University of Science and Technology (POSTECH), Pohang, 790-784, Korea}

\author{Yoon-Ho Kim}
\email{yoonho@postech.ac.kr}
\affiliation{Department of Physics, Pohang University of Science and Technology (POSTECH), Pohang, 790-784, Korea}

\date{\today}

%%%%%%%%%%%%%%%%%%%%%%%%%%
\begin{abstract}
We demonstrate the conditional reversal of a weak (partial-collapse) quantum measurement on a photonic qubit. The weak quantum measurement causes a nonunitary transformation of a qubit which is subsequently reversed to the original state after a successful reversing operation. Both the weak measurement and the reversal operation are implemented linear optically. The state recovery fidelity, determined by quantum process tomography, is shown to be over 94\% for partial-collapse strength up to 0.9. We also experimentally study information gain due to the weak measurement and discuss the role of the reversing operation as an information erasure.
\end{abstract}
%%%%%%%%%%%%%%%%%%%%%%%


\maketitle

%% Introduction %%

The projection postulate states that measurement of a variable of a quantum system irrevocably collapses the initial state to one of the eigenstates (corresponding to the measurement outcome) of the measurement operator and is one of the basic postulates of the standard quantum theory \cite{shankar,nie}. The initial state can never be recovered after a projection measurement on a quantum system.

If the measurement is not sharp (i.e., non-projective measurement), however, the situation is different. It is possible to reverse the measurement-induced state collapse  and the unsharpness of a measurement has been shown to be related to the probabilistic nature of the reversing operation which can serve as a probabilistic quantum error correction \cite{koashi}.  In particular, practical schemes for reversing the state collapse due to a weak (or partial-collapse) measurement in a solid-state qubit have been proposed in Ref.~\cite{koro} and one of the schemes has recently been demonstrated using a superconducting phase qubit in Ref.~\cite{katz}.  

Since single-photon states and linear optics play important roles in quantum communication and quantum computing research \cite{bouw,knill,gisin,kok}, it is of interest and importance to investigate how the measurement-induced state collapse due to a weak measurement can be reversed for a photonic qubit. In this letter, we report a linear optical implementation of conditional reversal of weak  (or partial-collapse) quantum measurements on a photonic qubit. We demonstrate experimentally that a nonunitary transformation of a photonic qubit, caused by a weak quantum measurement, can be reversed by applying an appropriately designed reversing operation. We also quantify and experimentally study information gain due to the weak measurement and discuss the role of the reversing operation as an information erasure.

%% Theory %%

Consider the initial state of a qubit represented in the computational basis,
%%%%%%%%%%%%%%%%
$
|\psi_{o}\rangle= \alpha |0\rangle+ \beta |1\rangle,
$
%%%%%%%%%%%%%%%%
where $|\alpha|^2 + |\beta|^2=1$. Ordinary projection measurement in the computational basis would collapse the state into  $|0\rangle$  (or $|1\rangle$) with the probability equal to $|\alpha|^2$ (or $|\beta|^2$). The projection measurement cannot be reversed because the projection operators 
%%%%%%%%%%
$\mathbb{P}_{0}=|0\rangle\langle
0|=\left(\begin{smallmatrix}
           1 & 0 \\
           0 & 0 \\
         \end{smallmatrix}\right)$
%%%%%%%%%%%%%
and 
%%%%%%%%%%%%   
$\mathbb{P}_{1}=|1\rangle\langle 1|=\left(\begin{smallmatrix}
           0 & 0 \\
           0 & 1 \\
         \end{smallmatrix}\right)$
%%%%%%%%%%%         
do not have mathematical inverse. An unsharp measurement on the qubit, however, can be reversible (although with a less than unity success probability) and the probability of successful reversal is related to the unsharpness of the measurement \cite{koashi}. 

The unsharp measurement that we consider in this paper is the weak or partial-collapse measurement discussed Ref.~\cite{koro,katz}, originally intended for a solid-state qubit. An essential part of the weak measurement is a detector, which measures the qubit, function as follows: the detector clicks with a probability $p$ if the qubit is in the $|1\rangle$ state and never clicks if the qubit is in the $|0\rangle$ state. The detector, thus, provides some partial information about the initial state of the qubit and, as we shall show later, the detector's output (click or no click) can be used to guess the initial state. 

Let us first assume that the detector has clicked. This situation is identical to the normal projection measurement 
 in which the state of the qubit is irrevocably collapsed to the $|1\rangle$ state. The measurement operator describing this situation can be written as 
%%%%%%%%%%%%%%%
$
M_{1}=\sqrt{p}|1\rangle\langle 1|=\begin{pmatrix}
                                    0 & 0 \\
                                    0 & \sqrt{p} \\
                                  \end{pmatrix}.
$
%%%%%%%%%%%
With no mathematical inverse, $M_1$ is not reversible and, therefore, is of no interest to us. 

Now, consider the situation in which the detector has not clicked.  The measurement operator $M_2$ corresponding to this situation can be evaluated by using the relation $\mathds{1}=M_{1}^{\dag}M_{1}+M_{2}^{\dag}M_{2}$ and is given by 
%%%%%%%%%%%%%
$
M_{2}=|0\rangle\langle 0|+\sqrt{1-p}|1\rangle\langle
1|=\begin{pmatrix}
   1 & 0 \\
   0 & \sqrt{1-p} \\
   \end{pmatrix}.
$
%%%%%%%%%%
The null output of the detector, therefore, corresponds to applying the $M_2$ measurement operator to the qubit and this is precisely the weak (or partial-collapse) measurement that we are interested in this paper.

The state of the qubit right after the null outcome of the detector is given as,
%%%%%%%%%%%
$
|\psi_m\rangle = {M_2 |\psi_o\rangle}/{\sqrt{\langle \psi_o| M^\dag_2 M_2 |\psi_o\rangle}} = \alpha'|0\rangle+ \beta'|1\rangle,
$
%%%%%%%%%%%%%%
where $\alpha' = \alpha/\sqrt{1-|\beta|^2 p}$ and $\beta' = \beta\sqrt{1-p}/\sqrt{1-|\beta|^2 p}$. To reverse the effect of the weak measurement, i.e., to recover the original state $|\psi_o\rangle$ from the state  $|\psi_m\rangle$, we only need to apply the inverse of $M_2$,
%%%%%%%%%%%%%%%
$
M_{2}^{-1}=\frac{1}{\sqrt{1-p}}\begin{pmatrix}
                                \sqrt{1-p} & 0 \\
                                         0 & 1 \\
                              \end{pmatrix}
$
%%%%%%%%%%%%%%%
, to the state $|\psi_m\rangle$ and the  reversing operation $M_2^{-1}$ exists mathematically as long as the variable $p$, defined as the partial-collapse strength, is less than unity. (The normal projection measurement corresponds to $p=1$.) 

Assuming that $M_2$ may be implemented for a photonic qubit, let us now examine how $M_2^{-1}$ can be realized experimentally. Since $M_2^{-1}$ can be re-written as  
%%%%%%%%%%%%%%%
$
M_{2}^{-1} = \frac{1}{\sqrt{1-p}}\begin{pmatrix}
                                 0 & 1 \\
                                 1 & 0 \\
                              \end{pmatrix}\begin{pmatrix}
                                             1 & 0 \\
                                             0 & \sqrt{1-p} \\
                                           \end{pmatrix}\begin{pmatrix}
                                                          0 & 1 \\
                                                          1 & 0 \\
                                                        \end{pmatrix} \equiv \frac{1}{\sqrt{1-p}} M_2^{\textrm{rev}},
$
%%%%%%%%%%%%%%%%
we define the physical implementation of the reversing operation as $M_2^{\textrm{rev}}$, i.e., the sequence of a bit-flip operation, another weak measurement $M_2$, and a final bit flip operation. The probability of successful reversal will always be less than unity and depend on the partial-collapse strength $p$ as $M_2^{\textrm{rev}}$ does not include the constant ${1}/{\sqrt{1-p}}$.

%% Experiment %%

\begin{figure}[t]
\includegraphics[width=3.0in]{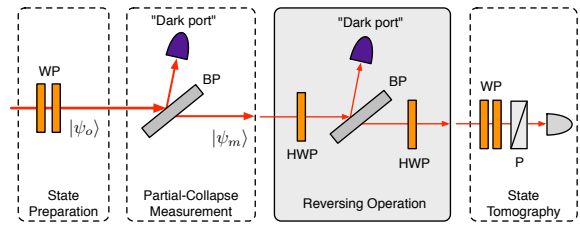}
\caption{Schematic of the experimental setup. The heralded single-photon state is used for encoding the polarization qubit $|\psi_o\rangle$ with a set of quarter and half-wave plates (WP). The partial-collapse measurement is implemented with a set of Brewster-angle glass plates (BP) and the partial-collapse strength $p$ is varied by increasing the number of BP's. The reversing operation requires two half-wave plates (HWP), in addition to the BP. Qubit state tomography is performed with WP and a polarizer (P).   }\label{setup}
\end{figure}

The experimental setup to implement the weak measurement and the reversal operation for a photonic qubit is schematically shown in Fig.~\ref{setup}. The single-photon state necessary for the implementation was prepared by spontaneous parametric down-conversion (SPDC) \cite{hong,baek}.  A 405 nm cw multi-mode diode laser was used to pump a 3 mm thick type-II BBO crystal to generate a pair of collinearly propagating SPDC photons centered at 780 nm (signal) and 842.4 nm (idler). The idler photon was detected at a trigger detector (not shown in Fig.~\ref{setup}) and an 80 nm bandpass filter was used in front of the trigger detector to reduce noise. When the trigger detector detects the idler photon, the signal photon is conditionally prepared in the single-photon state \cite{hong,baek}. The initial state of the photonic qubit $|\psi_o\rangle$ is then prepared by polarization encoding of the heralded single-photon state with a set of half-wave and quarter-wave plates (WP).

The weak (or partial-collapse) measurement on the photonic qubit is implemented by using an uncoated glass plate oriented at the Brewster angle (BP) and a single-photon detector positioned at the reflected mode, see Fig.~\ref{setup}. Since BP only reflects the vertical polarization state, $|1\rangle$, with a probability of reflection $p$, finding a  single-photon  in the reflected mode (identified by a click at the single-photon detector) is equivalent to subjecting the photonic qubit to $M_1$ measurement and this results in irreversible state collapse to the state $|1\rangle$. For the weak measurement, $M_2$, we must consider the conjugate outcome in which the reflected mode of BP is not occupied by the single-photon (hence the single-photon detector does not click). The null event at the single-photon detector in the reflected mode of BP (dark port in Fig.~\ref{setup}) unambiguously signals that the single-photon found in the transmitted mode of BP has been subjected to the weak measurement $M_2$ and the original state $|\psi_o\rangle$ has been partially collapsed to $|\psi_m\rangle$. 

%%%%%%%%%%%%%%%%%%%%%%%%%%%%%%%%%
%%%%%%%%%%%%%%%%%%%%%%%%%%%%%%%%%%

Since the scheme is implemented using the heralded single-photon state, there is no need to monitor the dark port in practice \cite{mitch}. Recording the coincidence event between the trigger detector and the detector placed in the transmitted mode of BP is sufficient to implement the weak measurement $M_2$ on the photonic qubit without any background noise.  The partial-collapse strength $p$ of the weak measurement can be increased by stacking more BP's and, in the experiment, $p$ is varied between 0.4 and 0.9 \cite{note}. The reversing operation $M_2^{\textrm{rev}}$ is implemented with two half-wave plates for the bit-flip operations and a set of BP's for the weak measurement $M_2$ whose partial collapse strength $p$ is identically set to the initial weak measurement.

Finally, coincidences between the trigger and the signal detectors were recorded and the signal detector was equipped with a 12.5 nm bandpass filter centered at 780 nm. To determine the state of the photonic qubit completely, quantum state tomography was performed to the heralded single photon state by making projection measurements in different measurement basis with WP and a polarizer (P) \cite{james}.

%% Quantum state tomography %%

In experiment, we tomographically analyzed the input state $|\psi_o\rangle$, the partially-collapsed state $|\psi_m\rangle$, and the recovered state by sequentially adding the partial-collapse measurement and the reversing operation, corresponding to a specific partial-collapse strength $p$, to the experimental setup for state preparation. The experiment was then repeated for a different value of $p$.

\begin{figure}[t]
\includegraphics[width=3.3in]{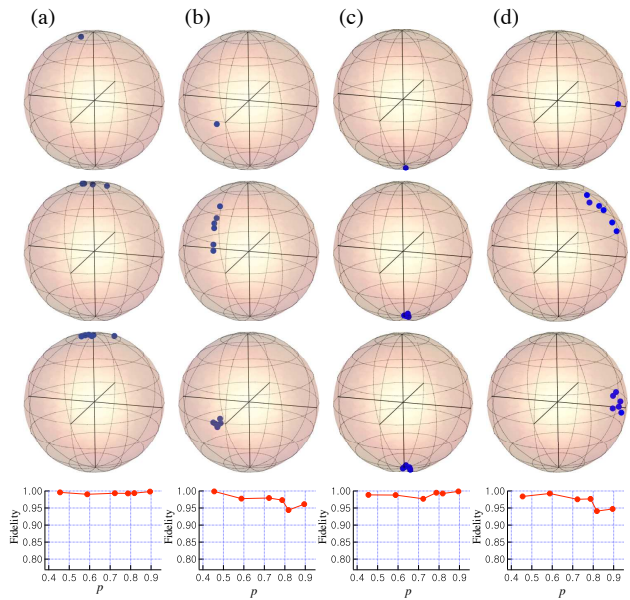}
\caption{The initial states (first row), the states after the partial-collapse measurement (second row), and the recovered states (third row) are represented on the Bloch sphere, as measured by quantum state tomography. For the second and the third rows, the points on the Bloch sphere correspond to varied partial-collapse strength $p$. The fourth row shows the fidelities between the initial states and the recovered states as functions of the partial-collapse strength. The input states are (a) $|H\rangle$, (b) $|L\rangle$, (c) $|V\rangle$, and (d) $|A\rangle$.  }\label{qst}
\end{figure}

Total of 14 input states were experimentally tested and, in Fig.~\ref{qst}, the results of quantum state tomography for four important input states  ($|H\rangle$, $|V\rangle$, $|A\rangle = (|H\rangle - |V\rangle)/\sqrt{2}$, and $|L\rangle = (|H\rangle + i |V\rangle)/\sqrt{2}$) are reported. It is evident from the experimental data that the partial-collapse measurement has little effect on the computational basis states $|H\rangle$ and $|V\rangle$, but the state collapse due to the measurement is clearly demonstrated for the other input states. The experimental data represented on the Bloch sphere also show that the reversing operation $M_2^{\textrm{rev}}$ restores the partially-collapsed state $|\psi_m\rangle$ back to the original state $|\psi_o\rangle$ quite faithfully: the fidelities (shown in the fourth row of Fig.~\ref{qst}) calculated between the recovered states  and the input states are shown to be over 94\% for all 14 input states and all the partial-collapse strength $p$ tested in the experiment.

%%%%%%%%%%%%%%%%%%%%%%%%%%%%%%%%%
%%%%%%%%%%%%%%%%%%%%%%%%%%%%%%%%%%

%% Quantum Process Tomography %%

The quantum state tomography data shown in Fig.~\ref{qst} for the four input states ($|H\rangle$,  $|L\rangle$, $|V\rangle$, and $|A\rangle$) then allow us to completely characterize quantum operations involved in this experiment, namely the weak measurement $M_2$ and the reversing operation $M_2^{\textrm{rev}}$ \cite{nie}. The matrix $\chi$, known as the quantum process tomography (QPT) matrix, completely characterizes the quantum operation and the QPT matrix $\chi$ was reconstructed experimentally from the experimentally reconstructed density matrices for the input states, the first row in Fig.~\ref{qst}, and for the recovered states, the third row in Fig.~\ref{qst}, using the maximum-likelihood estimation process \cite{fiu}.

\begin{figure}[t]
\centering
\includegraphics[width=3.0in]{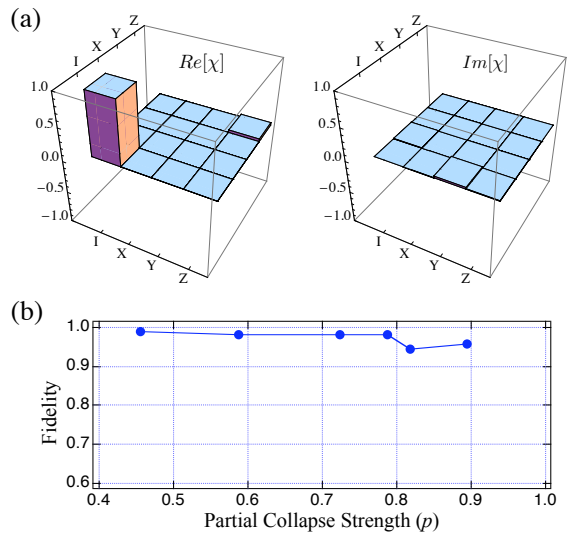}
\caption{(a) Quantum process tomography matrix $\chi$ for both the partial-collapse measurement and the reversing operation together at partial-collapse strength $p=0.895$. It is clear that the quantum process due to the partial-collapse measurement and the reversing operation together is mainly of the identity operation acting on the qubit. (b) The fidelity of the quantum process is over 94\% for all the partial-collapse strength $p$ tested in the experiment.}\label{qpt}
\end{figure}

In Fig.~\ref{qpt}(a), the QPT matrix $\chi$, in the Pauli matrix basis ($I, X,Y,Z$), for both the weak measurement $M_2$ and the recovering operation $M_2^{\textrm{rev}}$ together at partial-collapse strength $p=0.895$ is shown.   Since the reversing operation is supposed to completely (albeit probabilistically) recover the initial quantum state, the quantum operation involving both $M_2$ and $M_2^{\textrm{rev}}$ should ideally be an identity operation and the corresponding QPT matrix should be peaked at $(I,I)$ only for $Re[\chi]$. The experimental $\chi$ shown in Fig.~\ref{qpt}(a) clearly confirms this prediction. 

To have quantitative understanding on the performance of the reversing operation, we have determined the QPT matrices for both $M_2$ and $M_2^\textrm{rev}$ operations together for a number of partial-collapse strength $p$ and obtained the reversing fidelity $F=tr[\chi_{\textrm{exp}} \chi_{\textrm{ideal}}]$, defined as the overlap of the experimentally reconstructed QPT matrix $\chi_{\textrm{exp}}$ and the ideal one $\chi_{\textrm{ideal}}$ peaked at $(I,I)$ only for $Re[\chi_{\textrm{ideal}}]$. The result shown in Fig.~\ref{qpt}(b) demonstrates that the reversing operation functions quite well as designed: the fidelity of the quantum process is over 94\% for all partial-collapse strength tested in the experiment.

%%%%%%%%%%%%%%%%%%
%%%%%%%%%%%%%%%%%

%% Estimation fidelities.. and information gain %%

Let us now discuss the experiment in the context of information gain via the weak measurement. The detector output (click or no click) allows us to guess the initial state of the quantum system, $\rho_G$, and the quality of our guess can be quantified with the estimation fidelity, defined as $G_{\textrm{avg}}=\int \langle \psi_o| \rho_G | \psi_o \rangle d\psi_o$ \cite{bana,baek08}. 

First, consider the following guessing strategy: if the detector clicks (i.e., $M_1$ measurement occurs), we guess the initial state as $|1\rangle$ and if there is no click at the detector (i.e., $M_2$ measurement occurs), we guess that the qubit was more likely in the $|0\rangle$ state than in the $|1\rangle$ state \cite{koro,katz}. Therefore, our guess for the initial state of the qubit is $\rho_G^I = P_1 |1\rangle \langle 1| + P_2 \{p|0\rangle\langle0| + (1-p)|1\rangle\langle1|\}$, where $p$ is the partial-collapse strength, $P_1=\langle \psi_o| M_1^\dagger M_1|\psi_o\rangle=p|\beta|^2$, and $P_2=\langle \psi_o| M_2^\dagger M_2|\psi_o\rangle=|\alpha|^2 + (1-p)|\beta|^2$. It is then straightforward to show that $G_\textrm{avg}^I = (3+p^2)/6$. When no measurement is made (random guess; $p=0$),  $G_\textrm{avg}^I=1/2$ and if $p=1$ (projection measurement), $G_\textrm{avg}^I=2/3$, as it should be for a qubit \cite{bana,baek08}.

Although the above guessing strategy may look reasonable, it is not the optimal one. The optimal strategy is, in fact, to simply guess the initial state as $|0\rangle$ if the detector does not click ($M_2$ measurement). Formally speaking, the optimal guessing strategy must choose the eigenstate of a measurement operator associated with the largest eigenvalue \cite{bana}. For the $M_2$ operator, this corresponds to the $|0\rangle$ state. The guessed state is then $\rho_G^{II} = P_1 |1\rangle\langle1| + P_2 |0\rangle\langle0|$ and this leads to the estimation fidelity $G_\textrm{avg}^{II} = (3+p)/6$.

%%%%%%%%%%%%%%%%%%%%%%%%%%%%%%%%%
\begin{figure}[t]
\includegraphics[width=3.0in]{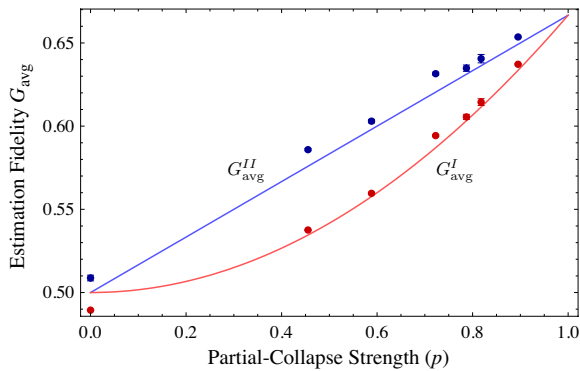}
\caption{Information gain via the weak measurement (quantified as $G_\textrm{avg}$) for the two guessing strategies discussed in the text as functions of the partial-collapse strength $p$. The random guess and the projection measurement correspond to $p=0$ and $p=1$, respectively. Solid lines are theory plots. }\label{info}
\end{figure}
%%%%%%%%%%%%%%%%%%%%%%%%%%%%%%%%%%

In Fig.~\ref{info}, we show the estimation fidelities for both guessing strategies. The experimental estimation fidelities are obtained by evaluating $\langle \psi_o| \rho_G | \psi_o \rangle$ from the experimental data for a set of initial states and then by averaging them \cite{baek08}. Clearly, the second strategy offers a better estimation fidelity when the partial-collapse strength does not correspond to the random guess ($p=0$) and the projection measurement ($p=1$).

The reversing operation then erases the information gained via the weak measurement. Since both the weak measurement and the reversing operation are successful only when all the reflected modes of BP's in Fig.~\ref{setup} are not occupied by a photon, the overall operation can be written as
%%%%%%%%%%%
$
M_2^\textrm{rev} M_2= \mathds{1}  \sqrt{1-p} 
$
%%%%%%%%%%%%%%
. Thus, the probability of successful reversal is $1-p$ regardless of the input state, indicating that the experimenter cannot learn anything about the initial state from the success probability determined with an identically prepared ensemble of qubits. The estimation fidelity in this case, obviously, is always 1/2 which is equivalent to randomly guessing the input state and, therefore, there is no information gain if the reversal operation is successful. In other words, the information gained via the weak measurement has been erased by the reversing operation.

%The guessing strategy after the reversing operation can be
%considered as below: if the first and the second detector clicks, we
%guess the initial state as $|1\rangle$ and $|0\rangle$,
%respectively. Since the second detector plays exactly opposite role
%to the first one, the strategy is convincing. When both detectors do
%not click (i.e, success reversingg case), however, there is no way
%to consider the initial state since the whole operation has the same
%amount of elements for $|1\rangle$ and $|0\rangle$. 

%%%%%%%%%%%%%%%%%%%%%%%%%%
In summary, we have demonstrated a linear optical implementation of the weak (partial-collapse) quantum measurement and conditional (probabilistic) reversal of the weak measurement for the photonic qubit. The quantum states and the involved quantum processes are quantitatively analyzed by using quantum state and process tomography techniques.  Moreover, we have quantified and experimentally studied information gain due to the weak measurement and discussed the role of the reversing operation as an information erasure.

YSK acknowledges the support of the Korea Research Foundation (KRF-2007-511-C00004). 
This work was supported, in part, by the Korea Science and
Engineering Foundation (R01-2006-000-10354-0) and the Korea Research
Foundation (KRF-2006-312-C00551).

%%%%%%%%%%%%%%%%%%%%%%%%%%%%%%%%%%%%%%


\begin{thebibliography}{}
\bibitem{note} The calculated and measured $p$ values for each BP set agreed well. For example, for two pieces of BP's, they are  0.478 and 0.455, respectively, and for seven pieces of BP's, they are 0.897 and 0.895.
\end{thebibliography}
\end{document}